# A Model for Web-Intelligence Index to Evaluate the Web Intelligence Capacity of Government Web Sites of Sri Lanka

**Prabath Chaminda Abeysiriwardana[1*] and S. R. Kodituwakku[2]**

[1]*Ministry of Science, Technology and Research, Colombo 03, Sri Lanka.*
[2]*Department of Statistics and Computer Science, Faculty of Science, University of Peradeniya, Sri Lanka.*

___________________________________________________________________________________

## Abstract

Web intelligence can be considered as a subset of Artificial Intelligence. It uses existing data in web to produce new data, knowledge and wisdom to support decision making and new predictions for web users. Artificial Intelligence is ever changing and evolving field of computer science and it is extensively used in wide array of web based business applications. Although it is used substantially in web based systems in developed countries, it is not examined whether it is being substantially used in Sri Lanka. Every Sri Lankan citizen depends on Public Service more or less throughout his/ her life time and at least more than 3 times: at birth, marriage and death. So providing most of these services to its citizen, Sri Lankan Government uses more or less of its country web portal. This paper presents a model to evaluate web intelligence capability based on weight to key functionalities with respect to web intelligence. The government websites were checked by the proposed criteria to show the potential of using web intelligent technology to provide website based services.

The result indicates that the use of web intelligence techniques openly and publicly to provide web based services through government web portal to its citizens is not satisfactory. It also indicates that lack of using the technologies pertaining to web intelligence in the public service web hinders the most of the advantages that citizen and government can gain from such technological involvement.

_______________________________________

*Corresponding author: E-mail: abeysiriwardana@yahoo.com;*

# 1 Introduction

## 1.1 Government Administration and Services

Sri Lanka is a multicultural country with many religions, ethnic groups and languages with population of 20.35 million as at 2012 census [1]. It is a democratic republic and a unitary state and the only country in South Asia that is rated "high" in 2014 on the Human Development Index [2]. For administrative purposes, Sri Lanka is divided into nine provinces and twenty five districts. The districts are further subdivided into 256 divisional secretariats, and there are approximately 14,008 Grama Niladhari divisions attached to these divisional secretariats. There are 3 types of government administration units in Sri Lanka namely districts, divisions and provinces. District and divisional level administration is done by central government while the provincial level administration is done by provincial government linked to the central government. The third and the lowest level administration linked to the central government is handled by three other types of local authorities that consist of 23 Municipal Councils, 41 Urban Councils and 271 Pradeshiya Sabhas. All of these administrative units have their own government institutes to provide public service free of charge or at discounted rate. Among these free services, there are few prominent public services like free education and free health service. Sri Lanka is one of the few countries in the world that provides universal free education from primary to tertiary stage [3] and free universal health care [4].

## 1.2 Data on Government Web

The data pertaining to the whole population dispersed throughout these institutes hasn't been centralized or decentralized properly. Most of the data are available in paper format or databases in isolated computers or computer systems. But government has taken initiatives to give more value to these data to provide more effective and efficient service to general public. They are mostly based on the Internet and web technologies.

One such initiative is known as Government Information Center of Sri Lanka. It provides information about 1439 services offered by 203 organizations. GovSMS links with Government Information Center to obtain about 13 services from government departments via Short Messaging Services (SMS). The GovSMS platform has SMS short code 1919, which is unified across all GSM and CDMA operators [5]. The official website of the Data and Information Unit of the Presidential Secretariat, Sri Lanka, has listed 148 government own web sites on its web site [6]. According to the government web portal, there are 656 websites pertaining to ministries, statutory bodies and corporations, provincial councils, district secretariats, divisional secretariats, departments, and government owned companies [7]. Most of the websites use two national/official languages namely Sinhala and Tamil [8], and the English as the link language to provide their services.

## 1.3 Use of Technologies to Disseminate the Government Information

Sri Lanka is the first country in the South Asian region to introduce 3G, 3.5G HSDPA, 3.75G HSUPA and 4G LTE mobile broadband Internet technologies [9] enabling these web services more accessible to its citizens. Despite this advancement in the computer science, Sri Lanka is far away from using new technologies effectively and efficiently in its public service through government websites. Although Sri Lanka advances to the $65^{th}$ place in 2015 from $76^{th}$ in 2014 of Network readiness index, it goes down to $92^{nd}$ position with respect to $2^{nd}$ pillar business and innovation environment of the same index [10]. One factor causing this is the availability of latest technology component ($70^{th}$ position) in the $2^{nd}$ pillar is not sufficiently satisfied by government owned websites which consist of a large part of all websites in Sri Lanka that influences this index.

## 1.4 Importance of Web-Intelligence in Government Web

One of the effective ways of handling huge amount of data pertaining to government web sites is to employ advanced technologies as much as possible. Web intelligence is one such technology in broad sense. Web intelligence is the research applications on the web having artificial intelligence incorporated in to information technology in order to create the next generation of products, services and frameworks which go beyond both artificial intelligence and information technology [11].

Today human needs, especially to be satisfied by the government in a postindustrial era, are so vast that information and services needed by its citizens are very much complex. Web-Intelligence is one such technology that can be used to satisfy these citizens' complex requirements effectively and efficiently. The human needs in a postindustrial era can be stated as below [12].

**1.4.1 Information empowerment**

Government has lot of information regarding its citizens such as information on birth, marriage, death and so forth. This information is inherently interrelated and used in different government services. If the information can be available in logical manner to the citizen and back again to the government information repository through the web, specifically through a web portal/ 'virtual organization', citizen is in position to consume the data very effectively and efficiently [13]. For example, in Sri Lanka to obtain birth certificates, citizens have to obtain a document from the hospital and then go to the registrar to fill another form with the same information. After some time they have to go to relevant authority to obtain the certificate. If there is a way to share these data by all necessary parties through web or through some intelligent agent, citizens only need to sit down before his/ her computer to obtain the relevant certificate through secured web system. This makes the life of authorities as well as citizens easy.

**1.4.2 Knowledge sharing**

Knowledge sharing is very prominent in modern education, health and agricultural practices in Sri Lanka. Student teacher interaction through web, doctors' advice to patients and community, especially in emergency cases and farmer agriculture instructors' interaction on crop related problems are ever increasing issues which can be effectively and efficiently addressed by knowledge sharing supported by web-intelligence techniques. For example, agriculturists and farmers need to interact with each other when problems arise in the field. If mobile applications are interrelated with web, farmers can post their problems through small phones so agriculturists can answer the problem at the office through web. Therefore, Web intelligence can facilitate the solving of problems encountered by farmers and agricultural instructors. Further the information on weather forecast and prediction of water levels of water resources such as tanks make available to the farmers through mobile applications as well as web using third party will facilitates the prediction and will make farmers life easy in the field.

**1.4.3 Virtual social communities**

Social communities are built to address special needs of citizens who need special attention. Social welfare programs guided by web-intelligence will address issues related to such people efficiently and effectively. There are lots of differently able people in Sri Lanka, who need special attentions. If societies are provided with information according to their needs, it would be great beneficial to them. For example, if government introduce web based job portal which can give expert advice in selecting proper jobs for these people, it will be even more effective in addressing the issues regarding these persons.

**1.4.4 Service enrichment**

This can be achieved through e-services available and popularized by using web-intelligence techniques.

**1.4.5 Practical wisdom development**

This is ultimate goal of web-intelligence and should be discussed when web-intelligence in government web reaches its maximum.

So it is very important to inspect the steps government has taken to initiate web-intelligence in its web sites and the effectiveness of the usage of technology. Therefore, web intelligence techniques can be used to enhance the quality of government services such as issuing birth/ death certificates, giving advice on government job seekers and so forth. It is important to carry out this survey to look at how effectively these techniques are used by the government to address these issues. It will give an overall idea about whether government has put enough weight on implementing these technologies in its web sites to achieve possible effective and efficient handling of citizens' requests. In this survey, government controlled 656 websites were tested to evaluate whether they are equipped with web-intelligence techniques or not. To test the web intelligence capacities on government web sites, basic study was carried out to define what kind of areas in web-intelligence should be tested on these government websites. After defining the areas of interests, a model based on a self-questionnaire was developed and applied to each web site for evaluating its web-intelligence capacity.

## 2 Materials and Methods

This section describes the materials used in this study and the overall research methodology.

### 2.1 Materials Used

In this study, 656 government controlled/owned websites relevant to ministries, statutory bodies and cooperation, provincial councils, district secretariats, divisional secretariats, departments, government owned companies were inspected during the months of March, April and May 2015.

So for this is the first research of this kind carried out in Sri Lankan public service. The 656 websites selected to inspect were listed under the Lanka Gate – the Country Portal (https://www.srilanka.lk) which is a gateway for citizens to access information and services provided by the government of Sri Lanka. All these web sites are under the same roof with so many similarities which could at the same time be considered as an advantage to improve WI capacity.

The validator software introduced in [14] is also used to test components which affect web-intelligence capacity of a web site. It should be noted that these software have not been developed specifically for testing web-intelligence capacity of a website directly but for testing some abilities of a website to reach web-intelligence capacity. This is important as most of the websites of the Sri Lankan government are in initial stage of web-intelligence capacities and they are not so competitive to be tested by advanced web-intelligence testing methodologies or tools.

### 2.2 Methodologies Used

The best methodology of testing whether a web site has web-intelligence is the inspection of coding and associated applications either hidden or prominent. Although it can be easily declared that there is some intelligence in a web site by testing some existing applications either by using black box test, behavioral test or similar kind of methodology, there is no concrete way to be sure of whether a web site has no intelligence in it when it doesn't have such publicly available web-intelligence application.

Checking the website for its intelligence is a kind of quality check of the website. Although it is not totally a quality check it checks an existing functionality associated with quality. Most of the existing evaluation methods assess the accessibility and usability of a website as indictors for quality and performance of website [15]. Usability test can be used to evaluate a website for its capacity of user computer interaction [16]. As web-intelligence increases the usability of a website, the proposed model asses the capacity of web-intelligence as an indicator to quality and performance of the website. Unfortunately there is no critical literature review or description regarding a survey or study regarding evaluation of web intelligence capacity on web sites which provides services to public through government intervention. Therefore a model is developed only to assess the presence of web-intelligence in a website and web-intelligence readiness of the website. So here only qualitative assessment plus simple quantitative assessment is developed through this model as an initiative to a more sensitive quantitative assessment which can be achieved through fuzzy logic and learning algorithms. Further it is assumed and justified that building a simple and not very sensitive model would be sufficient and appropriate to test web sites of government of Sri Lanka for its web-intelligence capacities.

The web sites were checked against whether public could access its intelligence services and if not websites were declared that they don't have web-intelligence capacity pertaining to public interest. The web-intelligence tested here is defined by presence of applications which have analytical power in data/ information handling, give new information/advise on citizens' questions, give instructions in orderly manner for particular request by citizens and presence of third party applications with same functionalities. Advance hidden applications specifically targeting web intelligence to achieve business intelligence were not considered as it is beyond this research. This can be justified since it explicitly says that the government websites are mainly for its citizens and public service. So black box testing or a kind of behavioral test (how it behaves – popup, expanding, new window with interacting facilities for certain actions – clicking, dragging, pointing) were combinely used to inspect these 600 plus web sites to check for their intelligence.

## 2.3 Questionnaires, Assumptions and Criteria

The criteria mentioned in the Table 1 were based on weight to key functionalities with respect to web intelligence within proposed model to evaluate web intelligence capability of web sites of Sri Lankan government. The government websites were checked by the proposed criteria to show the potential of using web intelligent technology to provide website based services. The criteria may directly support public interest in some instances.

The following two assumptions were made.

Assumptions:

1. The content in the government made or owned web site has no harmful or dubious or untruth information as they are controlled by government officials and standards of good governance.
2. A website reaches its web-intelligence acquired status by passing the no web-intelligence status and web intelligence ready status.

Thus the transition steps of a web-intelligence cooperated website were considered as follows.

No web-intelligence → web-intelligence ready → web-intelligence acquired

## 2.4 How to Calculate Web-Intelligence

Value column in Table 1 denotes the presence or non-presence status of a criteria (which has no sub criteria) and sub criteria (when a main criteria has sub criteria). So each and every websites were inspected to find out whether it supports or doesn't support the criteria/ sub criteria by giving value 1 or 0 against each and every criteria/ sub criteria. A criterion with sub criteria supposed to be satisfied when at least one of its sub criteria was satisfied. Weight is for measuring the impact of a criteria/ sub criteria on Web-Intelligence Index (WII). Once the criteria value set was completed, it was multiplied by Weight value set mentioned in Table 1. All criteria and sub criteria were considered to calculate the WII.

Thus WII was calculated by using Eq. (1),

$$WII = \Sigma V_i W_i \tag{1}$$

Although the sample of websites used in this research is huge it doesn't contain so highly competitive set of web sites to be evaluated for web-intelligence capacity. So the criteria were selected carefully to make only few lines of web-intelligence capacity distinguishable. So the values for Weight were carefully assigned arbitrary and relatively to each criteria to have fare impact of the criteria on the Web-Intelligence.

**Table 1. Criteria to measure web-intelligence capacity**

| Criteria no. (i) | Criteria | Value (V) | Weight (W) |
|---|---|---|---|
| 1 | Presence of the statements which indicates any web-intelligence capacity in the objectives of the web site | 0 or 1 | 0.05 |
| 2 | Presence of the technologies which supports intelligence in the website - Dynamic Languages like PHP, ASP, HTML5, Cloud computing | 0 or 1 | 0.15 |
| 3 | Presence of the Applications in the website which provides analytical power of data to user | | |
| | 3.1 Self created | 0 or 1 | 1.5 |
| | 3.2 Third party | 0 or 1 | 1 |
| 4 | Presence of the other documentation provided by the web site which states its web-intelligence capacity | 0 or 1 | 0.03 |
| 5 | By interacting with applications provided by the website - The services provided by the website like Online advertising, natural language understanding, personalized medicine, advising on education and etc. | 0 or 1 | 2.5 |
| 6 | By using software to check standards of the website. | 0 or 1 | 0.07 |
| | 6.1 W3C mobileOK Checker v1.4.2 - A Web Page is mobileOK when it passes all the tests [17] | 0 or 1 | 0.15 |
| | | 0 or 1 | 0.25 |
| | 6.2 Semantic Extractor - Sees a Web page from a semantic point of view [18]. - supports criteria 2 | 0 or 1 | 0.1 |
| | 6.3 RDF Validator - Checks and visualizes RDF documents [19]. – presence of a valid RDF document is enough | | |
| | 6.4 Feed Validator - Checks news feeds in formats like ATOM and RSS [20]. | | |
| 7 | By checking how the website stores and manages the incoming and outgoing data of it. | 0 or 1 | 0.15 |
| | 7.1 stores data on standard structure (database – SQL, Oracle, etc. / standard document format – XML, ontology, RDF). | 0 or 1 | 0.05 |
| | 7.2 stores in non-standard way (emails, text documents etc.) | | |

If 3rd or 5th criteria or both are satisfied the website provides web-intelligence capacity to its users and WII value is > 1

If at least one of 3rd or 5th criteria is not satisfied but any combination of 1st, 2nd, 4th, 6th, 7th criteria are satisfied website is considered as a web-intelligence ready website. Web-intelligence ready website is a web site which has potential to become a website with web-intelligence capacity. WII value ranging from 0 to 1 indicates that the site is in web-intelligence ready state.

If none of the above criteria are satisfied, website is destined as a website without any web-intelligence capacity or readiness for web-intelligence in its present status. The web-intelligence index value for this state is 0.

# 3 Results and Discussion

## 3.1 What is Specially Looked for?

When web intelligence is used in public service web sites it results in a combination of digital analysis, which examines how citizens consume public service and kind of government business intelligence, which allows a government's management to use data on citizens need, and demands trends to make effective strategic decisions. As the government institutes try to act as international level service provider, the need to analyze how citizens use its websites to learn about products and services is becoming increasingly critical to serve its citizens effectively and ultimate success [21].

The set of government web sites is not very competitive in web-intelligence capacity so basically the results were obtained looking for simple web-intelligence applications in these government websites as well as specific web-intelligence applications like online advertising, machine translation, natural language understanding, sentiment mining, personalized medicine, decision making in education and national security that are already in the public eye. At the same time some simple criteria were used to distinguish each type of WI and measure the depth of WI in these government web sites.

Lot of functionalities associated with these sites need a login using popular OAuth providers such as: Google, Facebook, LinkedIn etc. Examples of eServices offered by the government of Sri Lanka are certificates issuance, revenue license issuance & renewal, view results, tax payment, claim requests, bill payments, status requests and etc. These services range from checking the status of your eRevenue License to checking your results online. The eServices can be found categorized under government services and the A-Z Guide. It allows you to browse through all services provided by the Country Portal i.e. 26 e-Services, 13 mobile services (9 SMS services, 4 mobile apps) up to now [22].

## 3.2 Results Obtained and the Interpretation

All 656 government web sites were tested against the 7 main criteria and 8 sub criteria mentioned in materials and methods section and results obtained, are depicted in Tables 2 – 6.

Table 2 shows the number of sites that satisfied each criterion used in WII. The criteria 2 which states the presence of a suitable technology was met by more than 95% of sites. It could be seen that none of the site met the criteria of mobile OK or having a valid RDF document. It should be noted that even if RDF validator indicates the presence of RDF documents, it doesn't mean website has working web-intelligence capacity in it. This is because of the availability of RDF document to a WI application is not guaranteed by the RDF validator.

Only 1.5% sites stated in its objectives that it had incorporated web intelligence capacity to its website although according to the Table 4, 9.3% sites had WI capacity. 5.5% sites stated about its web intelligence capacity in some other place/ document of the site. This showed that government websites lack real insight into the web intelligence and its power regarding serving people more effectively and efficiently. It could be seen in Table 2 that high percentage of sites satisfied the presence of valid feed (criteria 6.4) and presence of standard structure (criteria 7.1) with vales of 71% and 89.5% respectively.

**Table 2. Number of web sites that satisfies each criteria**

| Criteria | 1 | 2 | 3 | | 4 | 5 | 6 | | | | 7 | |
|---|---|---|---|---|---|---|---|---|---|---|---|---|
| | | | 3.1 | 3.2 | | | 6.1 | 6.2 | 6.3 | 6.4 | 7.1 | 7.2 |
| No. of sites | 10 | 628 | 44 | 12 | 36 | 9 | 0 | 57 | 0 | 466 | 587 | 592 |
| Percentage of sites | 1.5 | 95.7 | 6.7 | 1.8 | 5.5 | 1.4 | 0.0 | 8.7 | 0.0 | 71.0 | 89.5 | 90.2 |

It could be seen according to Table 3 that more than one criterion had been satisfied simultaneously by most of the sites. Only 9 sites that is 1.3% of sites had satisfied single criteria only while 26 sites (4%) didn't satisfied any criteria. 94.7% sites had satisfied more than one criterion simultaneously. This is very good trend in achieving WI status from WI ready status and improving present WII value. According to the table 3, 61 sites (9.3%) achieved its WI status with at least more than one other WI ready component.

**Table 3. Number of web sites that satisfies each combination of criteria**

| | Criteria present | Result | No. of websites | No. of websites % |
|---|---|---|---|---|
| 1 | 1 only | WI ready | 0 | 0.0 |
| 2 | 2 only | WI ready | 7 | 1.1 |
| 3 | 3 only (if any from 2 components satisfied) | With WI | 0 | 0.0 |
| 4 | 4 only | WI ready | 0 | 0.0 |
| 5 | 5 only | With WI | 0 | 0.0 |
| 6 | 6 only (if any from 4 components satisfied) | WI ready | 1 | 0.1 |
| 7 | 7 only (if any from 2 components satisfied) | WI ready | 1 | 0.1 |
| 8 | 3 and 5 combination only | With WI | 0 | 0.0 |
| 9 | Any combination with (3 or 5) | With WI | 61 | 9.3 |
| 10 | Any combination excluding (3 or 5) | WI ready | 560 | 85.4 |
| 11 | Sites which has no combination | No WI | 26 | 4 |
| | Total | | 656 | 100 |

According to the Table 4, 61 sites (9.3%) had reached WI status while 86.7% sites were in WI ready state. Only 4% sites had no WI.

**Table 4. Total number of web sites that satisfied each category of WII**

| | With one criteria only | More than one criteria without WI criteria | WI criteria with WI ready criteria | More than one criteria of WI but with no WI ready criteria | Total | WI sites % |
|---|---|---|---|---|---|---|
| WI | 0 | 0 | 61 | 0 | 61 | 9.3 |
| WI ready | 9 | 560 | N/A | N/A | 569 | 86.7 |
| No WI | N/A | N/A | N/A | N/A | 26 | 4.0 |
| Total | | | | | 656 | 100 |

Table 5 shows 75.4% sites of total WI sites were at the lower range of WII value. These sites provided only some kind of analytical power of data to the users.

**Table 5. Total number of web sites against WII values**

| WII value | < 0.01 | 0.01 - 1.00 | 1.01 – 2.00 | 2.01 – 3.00 | 3.01 – 4.00 | 4.01 – 5.00 | 5.01-6.00 |
|---|---|---|---|---|---|---|---|
| No. of sites | 26 | 569 | 46 | 10 | 1 | 4 | 0 |
| % sites against total WI sites | | | 75.4 | 16.4 | 1.6 | 6.6 | 0.0 |

According to the Table 5, only 6.6% sites (4 sites) out of total WI sites went beyond the WII value of 4. These 4 sites are listed in Table 6 and provide some kind of decision making, advising, language understanding or some similar functionality to its web site users. It should be noted that in this evaluation only relative comparison is done so these sites got these high values through this model, only by means of presence of some applications which support web-intelligence capacity compared to other set of government web sites. The model facilitates this kind of comparison through its criteria and assigned weighted values. In measuring the quality of these applications, whether they are equipped with high quality techniques and critical web-intelligence capacities were not taken into account.

Table 6. Details of web sites that went beyond the WII value 4

| Name of the web site | Belongs to | Applications |
|---|---|---|
| www.crib.lk | Credit Information Bureau of Sri Lanka (CRIB) | The site provides analytical power of data to user through two login systems called CRIMS and STR. It has homemade applications as well as third party applications to interactively provide advising on financial matters. |
| www.hdfc.lk | Housing Development Finance Corporation Bank of Sri Lanka (HDFC) | The site provides online banking system which has analytical power of data to its users. It has homemade applications as well as third party applications such as online loan repayment system. These applications provide interacting capabilities to its users. |
| www.srilankajobs.net | Sri Lanka Job Bank | The site has applications which provide analytical power of data to user through its job bank. It has homemade applications as well as third party applications. It provides interacting with these applications to provide advising on job hunting. |
| www.youthjobs.lk | National Carrier Guidance and Counseling Center | The site has applications which provide carrier guidance and counseling to its users. It has homemade applications as well as third party applications such as carrier test and etc. These applications provide interacting capabilities to its users. |

One of the common characteristics of all the above four sites is that they store data and manage them in standard and non-standard ways.

It is worthwhile to note that there was no site which had satisfied all the main criteria at WII range 5.01-6.00.

## 3.3 What Government should Look for in the Direction of Web-Intelligence?

According to the final result it is apparent that although a lot of government web sites reach WI ready state (86.7%) only very few shows WI capacity (9.3%). This shows that government should take more steps to make this transition happen as the 21$^{st}$ century is the age of the Internet and the World Wide Web.

By the way of providing services, most of the web sites pertaining to government link largely with the areas of interest, such as agriculture, education, food security, health, transport and social welfare. Each area links with some kind of business activities such as payment processing for the service provider by the relevant authority. It is observed that ecommerce facilities exist for some sites, but lacks the interconnectivity between the services followed by. For example, if you want to renew your revenue license online, the order of payment for insurance, echo test and so forth, is important and users are in need of connecting more than two websites and more than two services. So in order to provide better services to the citizen, existing of a common agent is necessary to govern the process and provide guidance to website users who are not aware of how to start and continue the process.

In this research, some dynamic nature in orientation of government services within the domain of web portal was observed. Within the short evaluation period, some services provided by some web sites were disappeared and while some of them were reappeared again in some other websites. While the total number of services being intact, this behavior might be due to the change of functionalities assigned to governing authorities of these websites by the government. The bad side of such change is that some of these changes were taking place without notice to the web site users. This kind of issues could be handled by using different strategies. One such strategy would be the use of intelligent agent assigned to the web portal to look after whole set of services and once such change happens make a flag on respective web sites and their administrators about the change and its consequences. This would introduce prompt remedial actions to correct the problem that might be faced by unaware web site users.

Most advance web sites use internal web intelligence related applications for data mining which are not in the reach of the surfing site users and highly instrumental in providing most satisfactory services to its users. According to the analysis of overall result of this evaluation, the government web sites significantly lack such kind of internal applications. For example, government asks all its state banks not to issue Automatic Teller Machine (ATM) cards to government pensioners who receive pension after their retirement from government service due to risk of retrieving the pension by any other illegal person after the death of pensioner. This is not a good practice as it hinders the ability of providing this good social welfare program effectively to this special society while the technologies to handle such issues are available. If technologies pertaining to web intelligence are supposed to be applied to curb this problem, the process can be started at the point when the doctor confirms the death of the pensioner at the hospital. Initial data pertaining to death certificate can be stored in a hospital database in a logical manner using any of RDF, XML, ontology or such structured document. The web mining technologies in state bank could be used to process such data while intelligent agent in a central database of government web portal which stores pensioners' personal data, is monitoring and facilitating the process of updating of the same at every other location. This will be used to immediately stop the illegal retrieval of money from pensioner's bank account by ATM card after the death of pensioner. Another solution could be based on data mining from a social web forum formed by pensioners. Anyway this lack of basic data mining techniques in government web should be further studied and addressed properly to make them sufficiently available in government web to resolve such kind of issues.

The government should take advantage of the web which is in the process of redefining the meanings and processes of business, commerce, marketing, finance, publishing, education, research and development [23]. Although individual web based information systems are constantly being deployed by government web sites (89.5%), advanced issues and techniques for developing and for benefiting from web intelligence still remain to be a huge challenge in the way of integrating such individual web based information systems to one public web portal.

Especially when education, health and agriculture fields are considered, substantial amount of websites pertaining to them can be found. The amount of data pertaining to them is very large. The number of applications associated with them is also large. People interactions with them seem very high. But data were scattered between web pages and applications. If there were logical interactions between these application and data the result would have been highly effective in solving citizens need. It would be easy to identify how peoples need should be approached if there was a method to analyze statistics regarding people's interaction with these websites. When considering the amount of freely available third party web intelligence software, the amount used in government web was not in satisfactory level although in some cases the third party web-intelligence software were used.

Any way Sri Lankan government already has good web portal which could be improved in to an Intelligent Portal. It would be an intelligent Virtual Industry Park (VIP) a kind of enterprise portal discussed by Ning Zhong [13]. So the co-evolution of government web with web-intelligence will one day result in more user friendly and efficient government web services to its citizens [24]

## 4 Conclusion

In this paper, the importance of WI in modern web development is discussed. If government web sites can incorporate these WI capacities to its web sites, government will be in good position to provide good government service to its citizens through its web sites.

In this research, 656 government web sites were checked for web-intelligence capacity. It was found less than 10% sites have WI. Also it was found that more than 80% sites have capacity to reach WI status. The model used to evaluate the government web site use simple arbitrator and weighted values as the evaluated web sites are not so competitive and sensitive to the web-intelligence capacities. But it is recommended to use fuzzy and learning algorithms to improve this model if it is used in more competitive and sensitive set of web-intelligence web sites.

It is observed that the use of web intelligence techniques open and publicly to provide web based 355 services through government web portals to the citizens are not satisfied. So the lack of using the technologies pertaining to web intelligence in the public service web hinders the most of the advantages that citizen and government can gain from such technological involvement.

## Competing Interests

Authors have declared that no competing interests exist.

## References

1. Statistics of Sri Lanka - Department of Census and Statistics Sri Lanka.
   Available: http://www.statistics.gov.lk/ (Accessed 20 April 2015).

2. 2014 Human development report summary. United Nations Development Programme. 2014;21–25.
   Available: hdr.undp.org/sites/default/files/hdr14-report-en-1.pdf (Accessed 27 April 2015).

3. Education of Sri Lanka.
   Available: http://countrystudies.us/sri-lanka/46.htm (Accessed 20 April 2015).

4. Work in Sri Lanka.
   Available: http://workinsrilanka.lk/living/healthcare-in-sri-lanka (Accessed 20 April 2015).

5. Government Information Center of Sri Lanka.
   Available: http://www.gic.gov.lk/gic (Accessed 15 April 2015).

6. Government Websites, the Official Website of the Data and Information Unit of the Presidential Secretariat, Sri Lanka.
   Available: http://www.priu.gov.lk/Gov_Links.html (Accessed 14 April 2015).

7. A-Z Gov Web Index.
   Available: http://www.gov.lk/ (Accessed 14 April 2015).

8. Official Languages Policy, Department of Official Languages.
   Available: http://www.languagesdept.gov.lk/ (Accessed 10 April 2015).

9. Cellular Services, the Telecommunications Regulatory Commission of Sri Lanka.
   Available: http://www.trc.gov.lk/cellularservices.html (Accessed 15 April 2015).

10. Network Readiness Index, Global Information Technology Report 2015- The World Economic Forum.
    Available: http://reports.weforum.org/global-information-technology-report-2015/network-readiness-index (Accessed 17 April 2015).

11. Web intelligence, Wikipedia, the free encyclopedia.
    Available: http://en.wikipedia.org/wiki/Web_intelligence (Accessed 13 April 2015).

12. Liu J. Web intelligence (WI): what makes wisdom web? Proceedings of Eighteenth International Joint Conference on Artificial Intelligence (IJCAI-03). 2003;1596–1601.

13. Zhong N, Liu J, Yao YY. "Web Intelligence (WI)", the Encyclopedia of Computer Science and Engineering. Wiley. 2009;5:3062-3072.

14. Browse W3C's Open Source Software.
    Available: http://www.w3.org/Status.html (Accessed 16 April 2015).

15. Kavindra Kumar Singh, Praveen Kumar. A model for website quality evaluation - a practical approach. International Journal of Research in Engineering & Technology. 2014;2(3):61-68.

16. Soohyung Joo, Suyu Lin, Kun Lu. A usability evaluation model for academic library websites: Efficiency, Effectiveness and Learnability. Journal of Library and Information Studies. 2011;9(2):11-26.

17. W3C mobileOK Checker, W3C.
    Available: http://validator.w3.org/mobile (Accessed 10 April 2015 - 28 May 2015).

18. Semantic Data Extractor, W3C.
    Available: http://www.w3.org/2003/12/semantic-extractor.html (Accessed 10 April 2015-28 May 2015).

19. Check and visualize your RDF documents, W3C.
    Available: http://www.w3.org/RDF/Validator (Accessed 10 April 2015 - 28 May 2015).

20. Feed Validation Service, W3C.
    Available: http://validator.w3.org/feed (Accessed on 10 April 2015 - 28 May 2015).

21. Overview, Web Intelligence, UCI Extension.
    Available: http://unex.uci.edu/areas/it/web_intel/ (Accessed 16 April 2015 - 28 May 2015).

22. Lanka Gate – Country Portal.
    Available: https://www.srilanka.lk (Accessed 10 April 2015).

23. Zhong Ning, Liu Yao, Jiming Yao YY, Ohsuga S. "Web Intelligence (WI)", Web Intelligence, Computer Software and Applications Conference, 2000. COMPSAC 2000. The 24th Annual International. 2000;469.
    DOI: 10.1109/CMPSAC.2000.884768, ISBN 0-7695-0792-1

24. Yao YY. Web-based research support systems. Proceedings of the second international workshop on web-based support systems (WSS 2004), Yao JT, Raghavan VV, Wang GY (Eds.), Beijing, China, 2004;1-6.